# Orbital solution and evolutionary state for the eclipsing binary 1SWASP J080150.03+471433.8

M. S. Darwish[1,2], M. M. Elkhateeb[1,3], M. I. Nouh[1,3*], S. M. Saad[1,2], M. A. Hamdy[1,2],

M. M. Beheary[4], K. Gadallah[4] and I. Zaid[1,3]

Email: abdo_nouh@hotmail.com

Tel: +20 2 25543111
Fax: +202 25548020

[1] Astronomy Dept., National Research Institute of Astronomy and Geophysics, 11421, Box:138, Helwan, Cairo, Egypt.
[2] Kottamia Center of Scientific Excellence in Astronomy and Space Sciences (KCScE).
[3] Physics Dept., College of Science, Northern Border University, 1320, Arar, Saudi Arabia
[4] Astronomy Dept. and Meteorology, Faculty of Science, Al-Azhar University, Cairo, Egypt

**Abstract:** We present an orbital solution study for the newly discovered system 1SWASP J080150.03+471433.8 by means of new CCD observations in VRI bands. Our observations were carried out on 25 Feb. 2013 using the Kottamia optical telescope at NRIAG, Egypt. 12 new times of minima were estimated and the observed light curves were analysed using the Wilson-Devinney code. The accepted orbital solution reveals that the primary component of is more massive and hotter than the secondary one by about $28^0$K. The system is an over-contact one with fillout ratio ~ 29% and is located at a distance of 195 Pc. The evolutionary status of the system is investigated by means of stellar models and empirical data.

## 1. Introduction

The study of short period eclipsing binaries is important for understanding the nature and evolution of low-mass stars and to allow investigation of the cause of the period cut-off (Norton et al, 2011). The system 1SWASP J080150.03+471433.8 (we will further use the short name SWAP08) is one of 53 short period eclipsing binary stars identified by Super WASP project (Norton et al, 2011). The identified list of stars includes 48 new objects with periods < $0.^d23$. The system SWAP08 was classified as a short period W UMa star (p=$0.^d21751$) with $V_{max}$ = 13.40 mag, while the depth of the primary and secondary minima are 0.66 and 0.64 mag respectively. The first photometric observations were obtained for the system after its discovering by Terrill and Gross (2014) during their reclassification for 143 new W UMa systems identified by Lohr et al. (2013) in the SuperWasp data. They measured complete light curves in the B, V, and $I_c$ bands and established the first photometric solution for the system.

A photometric study of the eclipsing binary SWAP08 was our goal using new VRI CCD observations. The present paper is a continuation of the series concerning the photometric analysis of the newly discovered eclipsing binaries, Elkhateeb and Nouh (2016), Elkhateeb et al. (2015), Elkhateeb et al. (2015), Elkhateeb and Nouh (2015) and Elkhateeb et al.(2014)

Section 2 is devoted to the observations. In section 3, a light-curve analysis of the system is given. Section 4 deals with the evolutionary state of the two components and section 5 is the conclusions.

## 2. Observations

New CCD observations were carried out in the VRI bands on 25 February 2013 using a 2Kx2K CCD camera attached to the 1.8 m Kottamia optical telescope. Differential photometry was performed with respect to GSC 3408-01475 and GSC 3408-00253, as comparison and check stars, respectively.

A total of 147 measurements were obtained. The corresponding phases of the observed data were calculated using the ephemeris adopted from our observations as:

$$\text{Min I} = 2456714.3010 + 0.217513 * E \quad (1)$$

Table 1 lists the magnitude difference (the variable minus the comparison star) in VRI, together with the corresponding Julian date and phase, and these measurements are displayed in Fig. 1. The parameters describing the observed light curves **$D_{max}$ = (Max I – MaxII) and $D_{min}$ = (Min I – Min II)** have been measured and listed in Table 2 together with the depth of the primary (**$A_p$ = (Min I – Max I)**) and the secondary (As = (Min II – Max I)) minima in the VRI bands. Using the method of Kwee and Van Woerden (1956) a total of 12 new times of primary and secondary minima (six minima estimated from our observations and another six from Terrill and Gross (2014) light curves) were derived by means of the software Minima V2.3 (Nelson 2002) and are listed in Table 2.

## 3. Photometric Analysis

As the light curves of the system, SWAP08 showed deep potentially complete (total/annular ), eclipses, it was selected by Terrell and Gross (2014) for B, V, and $I_c$ observations. They carried out a first photometric study and a set of light curve parameters were estimated using a W-D code. They fixed the mean surface temperature of star 1 ($T_1$) at 4500 $^0K$ based on the APASS (B-V) value and the relatively low interstellar reddening in the field. Terrell and Gross suggest in their conclusion to restudy the system SWAP08 using larger instruments.

Terrell and Gross (2014) noted that the observed light curves for the system SWAP08 show a mild asymmetry between the two maxima in SuperWASP observations which didn't display in their observation. They suggest that may refer to the presence of spot phenomena with a variable in size/or location.

In the present paper, we try to contribute some results that may be added or correct any estimated results by Terrell and Gross based on a photometric analysis for the observed light curves in our VRI observations. Our observations show smooth light curves especially in the two maxima and minima. These regions show some scattering in SuperWASP and Terrell and Gross observations which may have affected their photometric solutions. We applied the photometric parameters estimated by Terrill and Gross to our normalized light curve in the VRI bands which showed that they did not match well, especially in the two maxima of the light curves and the depth of the secondary minimum.

We used the adopted photometric parameters by Terrill and Gross (2014) as initial values through a 2009 version of the Wilson and Devinny code (Nelson 2009), which is based on model atmospheres by Kurucz (1993). Gravity darkening and bolometric albedo exponents for convective envelopes, were $A_1 = A_2 = 0.5$ (Rucinski, 1969) and $g1 = g2 = 0.32$ (Lucy 1967). Bolometric limb darkening was adopted using tables of Van Hamme (1993) using the logarithmic law for the extinction coefficients. Modifications to some initial values of the parameters (e.g. temperature of the primary component $T_1= 4589$) were made in order to obtain a good matching between the normalized and theoretical curves.  The adjustable parameters for the light curve solution are the orbital inclination (i), the mass ratio (q), the temperature of the secondary component ($T_2$) and surface potential ($\Omega$). The condition of Mode 3 (over contact mode) in the W-D code was applied and the best photometric fitting was reached after several runs. The estimated parameters (see Table 4) show that the primary component is hotter and more massive than the secondary one, while the temperature difference between the components is only about $28^0$K. Figure 2 displays the observed light curves together with the synthetic curves in VRI.

The spectral type of both components of the system SWAP08 was adopted as K4 (Popper, 1980) according to the parameters of the accepted orbital solution.

The three-dimensional geometric structure of the system SWASP08 was constructed using the Binary Maker 3.03 software Package (Bradstreet and Steelman, 2002) based on the calculated parameters and is displayed in Figure 3.

The empirical $T_{eff}$ − mass relation by Harmanec was used to estimate the absolute physical parameters for the components. The resulting mass of the primary component is $M_1(M_\odot) = 0.70 \pm 0.29$ while the mass of the secondary component was calculated directly from the estimated mass ratio of the system ($q = M_1/M_2$) as $M_2(M_\odot) = 0.34 \pm 0.01$.

The radii $R_1$ ($R_\odot$) and $R_2$ ($R_\odot$) of the primary and secondary components, respectively were calculated and listed in Table 5. The distance (d) to the system SWAP08 was calculated by means of the estimated photometric and absolute parameters ($d = 10^{(m-M_v+5)/5}$), where m and $M_v$ are the apparent and absolute magnitude for the studied system, respectively. The calculations showed that the average distance is 195 pc.

## 4. Discussion and Conclusion

New VRI CCD observations for the new eclipsing binary system SWASP08 were carried out and new times of minima were calculated in each filter using the Kwee and Van Woerden (1956) method. A new orbital solution was obtained by analyzing the observed light curves by means of the Wilson-Divenny code (Nelson 2009). The primary component is more massive and hotter by $\Delta T \approx 28^0 K$ than the secondary, and both belong to spectral type K4. The Geometric configuration shows that SWASP08 is an over contact binary with fill out ratio ~29%. The calculations led to an average distance of 195 Pc. A set of physical parameters was estimated which places the components of the system in the HR diagram.

The computed physical parameters listed in Table 5 are used to investigate the evolutionary state of the systems. For this purpose we used the evolutionary tracks computed by Girardi et al. (2000) for both zero age main sequence stars (ZAMS) and terminal age main sequence stars (TAMS) with metalicity Z = 0.019. The locations of the two components on the M-R and M-L diagrams are shown in Figure 4 and Figure 5. The behaviour of the system compatible with the short period binaries that the primary component lies on the ZAMS while the secondary component lies above the TAMS.

To compare the physical parameters with the empirical data, we used the most recent mass-radius relation by Eker et al. (2014) which consists of similar well-known systems, and the mass-luminosity relation (MLR) of Eker et al. (2015). Figures 6 and 7 illustrate these M-R and M-L relations. The primary component of the system showes a good fit with the empirical M-R relation, while the secondary component lies above it as is the case in a quite number of contact binaries of this type. The deviating position of the secondaries is ascribed to energy transfer from primary to secondary through the common convective envelope, as suggested by Lucy (1973). These results are similar to that of the evolutionary tracks, Figures 4 and 5.

**Acknowledgments:** This research has made use of NASA's Astrophysics Data System. This research is supported by grants by Science and Technology Development Fund (STDF) N5217. We are very grateful to the team of Kotamia Astronomical Observatory, Dr. F. I. Elnagahy, M. Ismail, D. El Sayed and I. Helmy. We would like to give thanks to Dr. N. M. Ahmed, D. Fouda, A.Shokry and M. Eldepsy for their helpful discussions and advice. The authors would to thank the referee for his/her valuable comments.

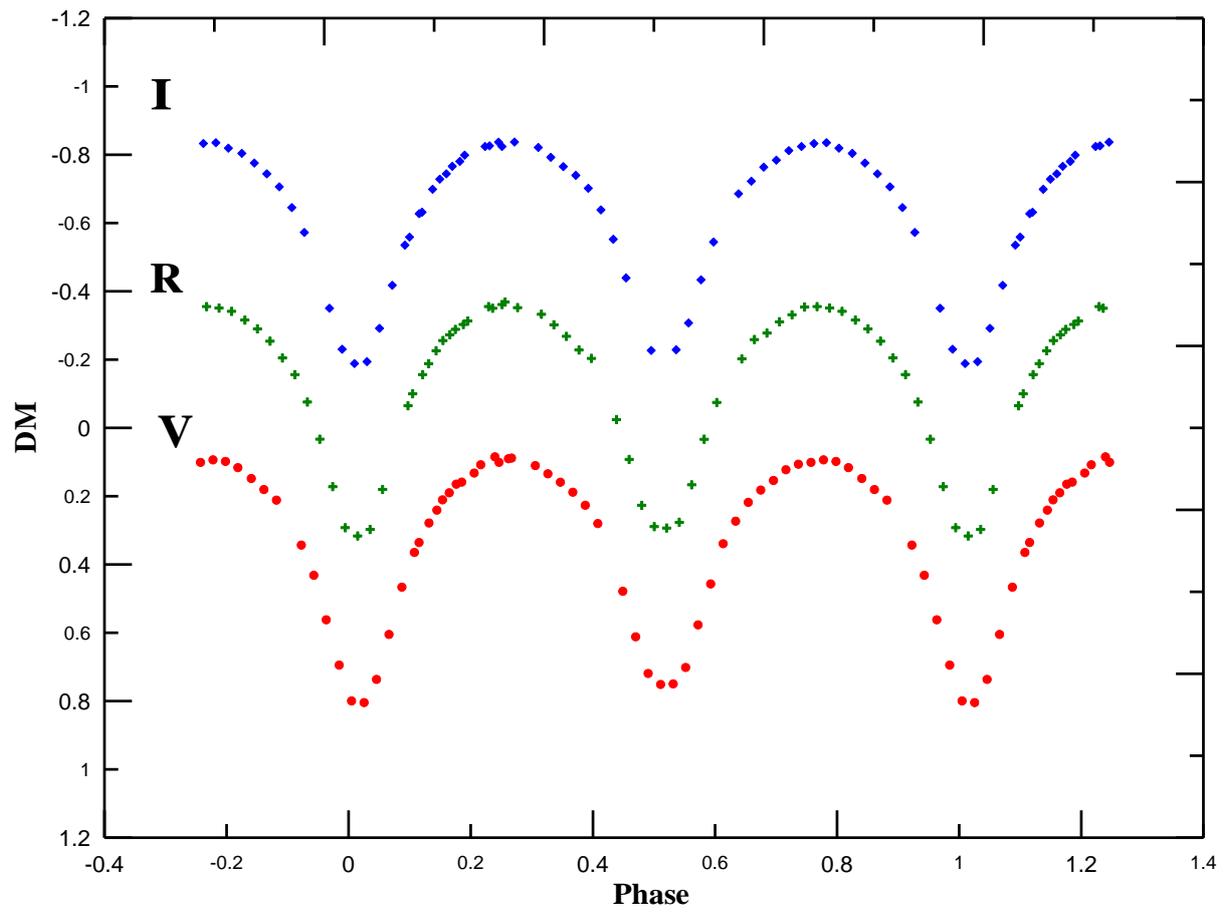

**Fig.** 1. Observed light curves of the system SWASP08 in VRI bands

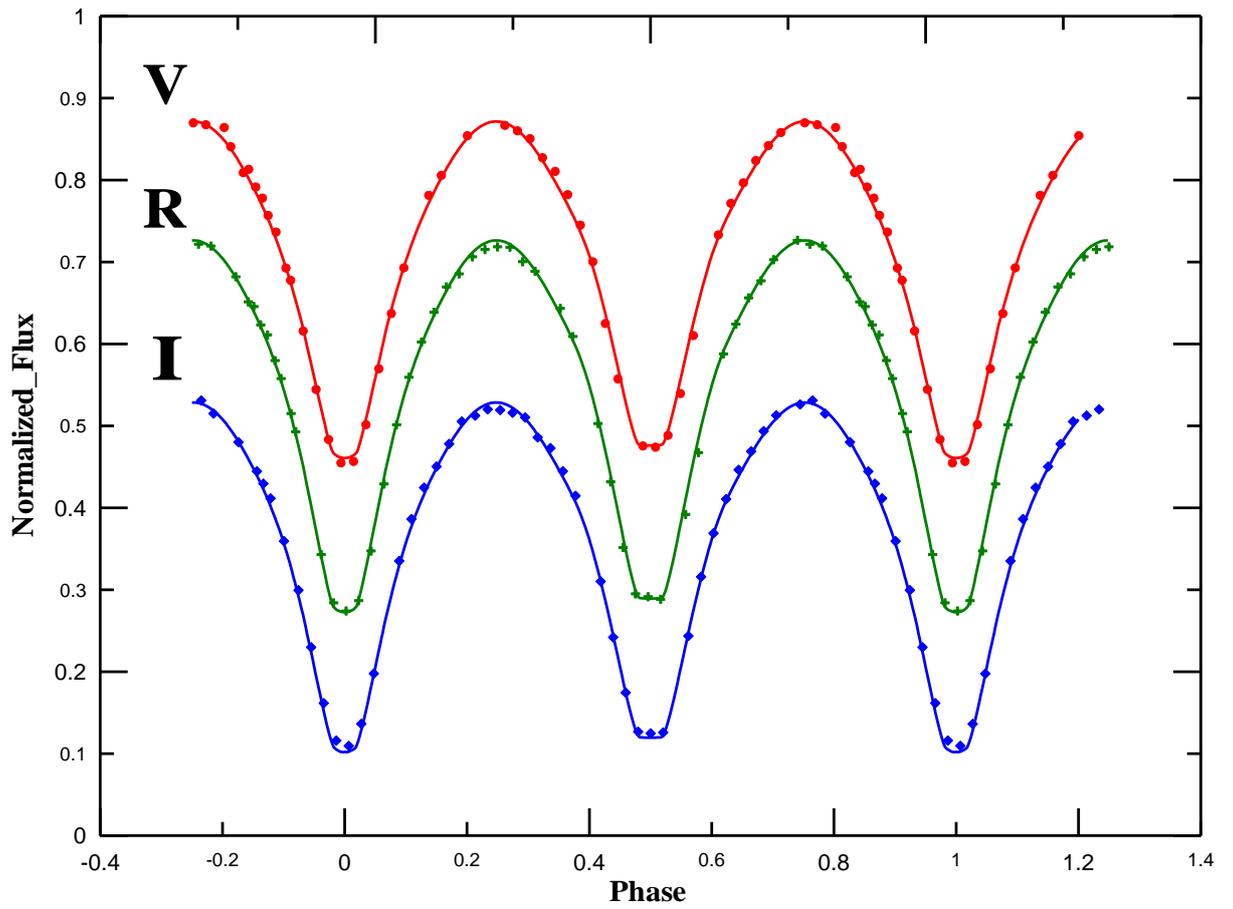

**Fig.**2. Observed light curves (filled circles), and the synthetic curves (solid line) in VRI bands.

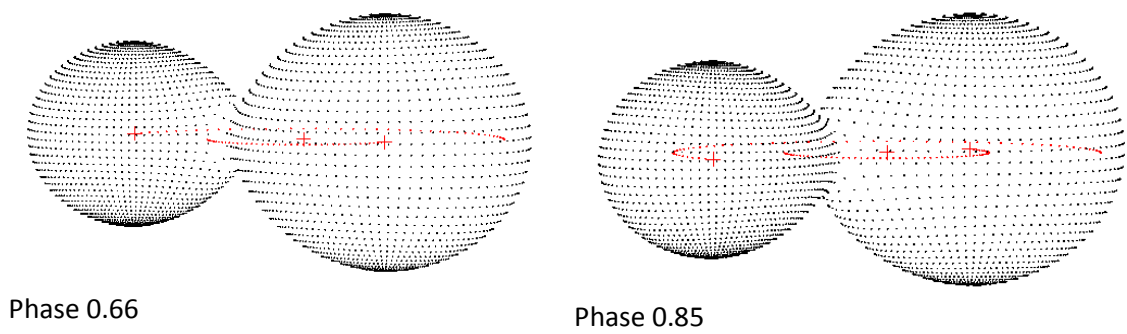

Phase 0.66　　　　　　　　　　Phase 0.85

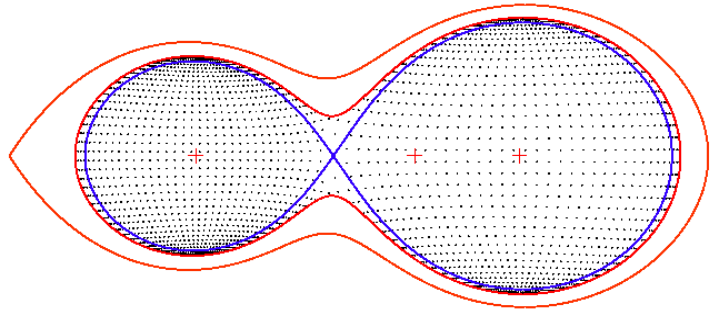

**Fig. 3.** Geometric structure of the binary system SWASP08.

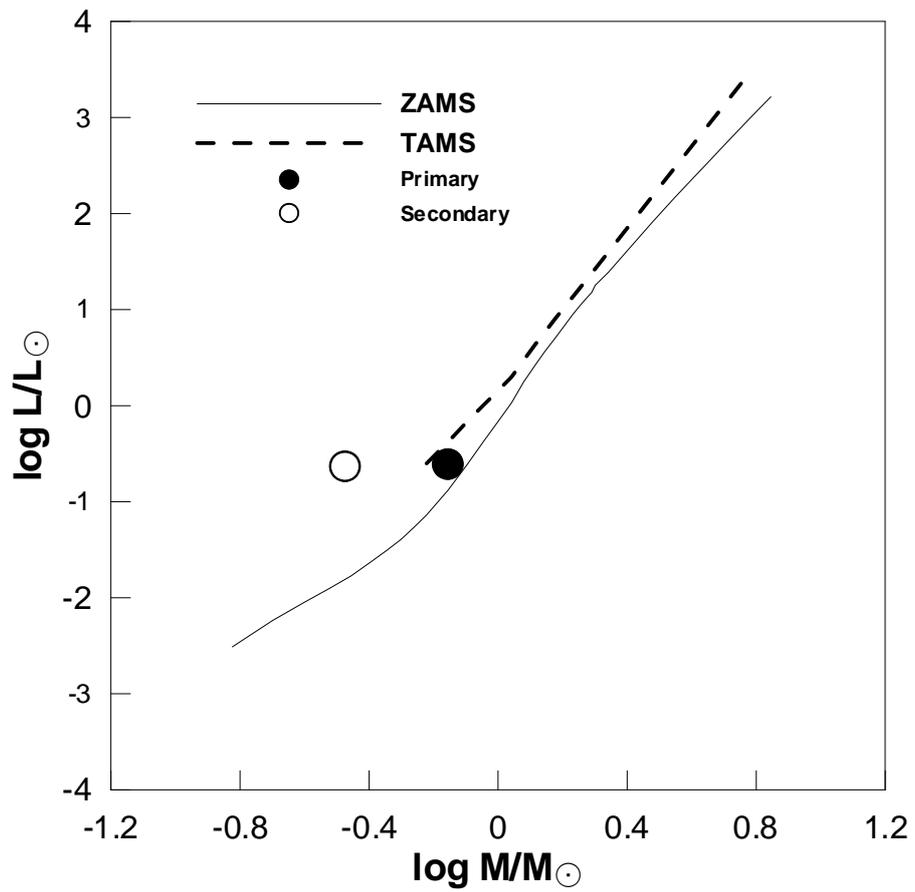

Figure 4: Positions of the two stars of SWASP08 on the theoretical mass–luminosity diagram of Girardi et al. (2000).

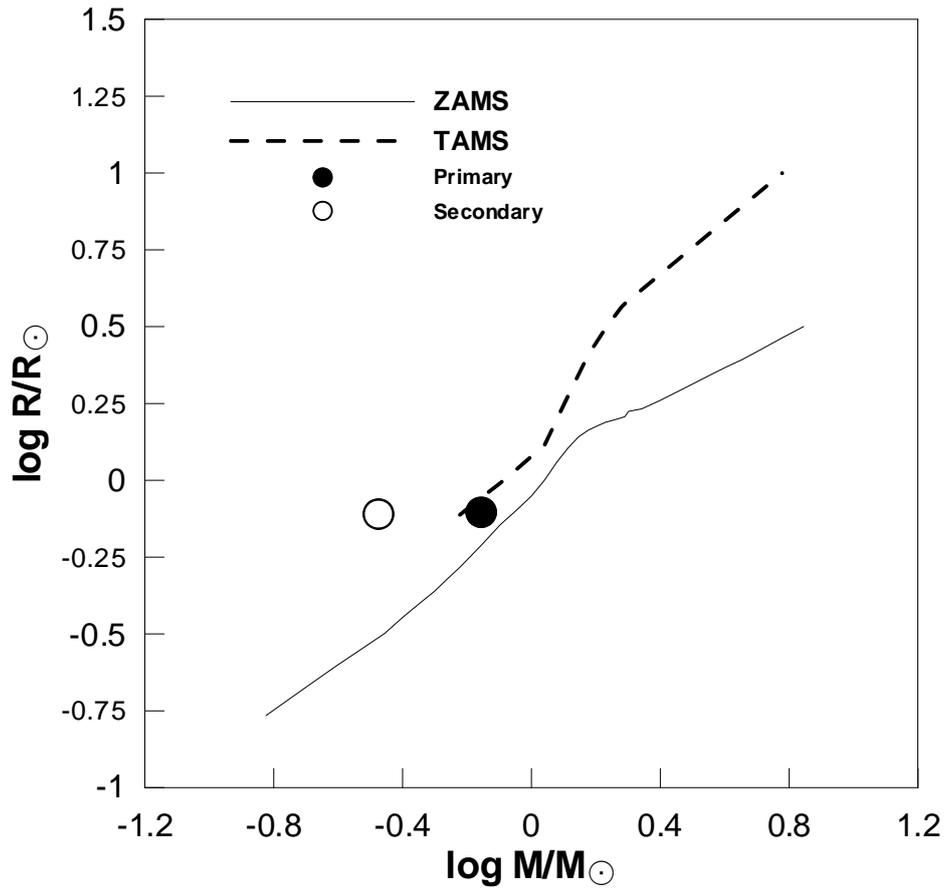

Figure 5: Positions of the two stars of SWASP08 on the theoretical mass–radius diagram of Girardi et al. (2000).

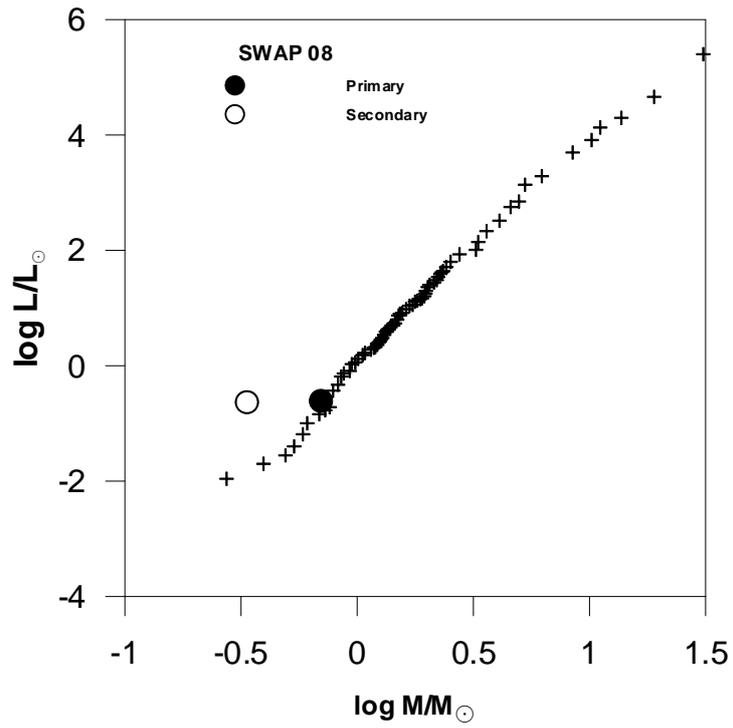

Figure 6: Location of the two stars of SWASP08 on the empirical M-L diagram of Eker et al. (2014).

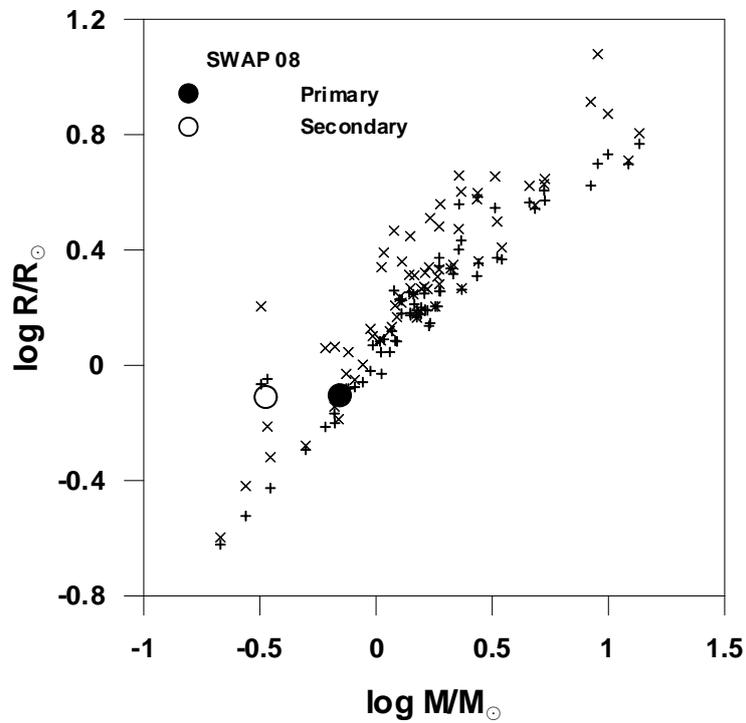

Figure 7: Location of the binary SWASP08 on the empirical M-R diagram of Eker et al. (2014).

Table 1. VRI observational data of the eclipsing binary SWASP08.

| JD | ΔV | Error | Phase | JD | ΔR | Error | Phase | JD | ΔI | Error | Phase |
|---|---|---|---|---|---|---|---|---|---|---|---|
| 2456714.2443 | 0.0883 | 0.0034 | 0.2605 | 2456714.2466 | -0.3634 | 0.0028 | 0.2501 | 2456714.2478 | -0.8387 | 0.0029 | 0.2444 |
| 2456714.2491 | 0.0829 | 0.0034 | 0.2384 | 2456714.2514 | -0.3573 | 0.0028 | 0.2280 | 2456714.2526 | -0.8261 | 0.0029 | 0.2223 |
| 2456714.2542 | 0.1059 | 0.0031 | 0.2152 | 2456714.2604 | -0.3050 | 0.0029 | 0.1866 | 2456714.2616 | -0.7824 | 0.0030 | 0.1810 |
| 2456714.2629 | 0.1628 | 0.0036 | 0.1751 | 2456714.2652 | -0.2747 | 0.0029 | 0.1646 | 2456714.2664 | -0.7462 | 0.0030 | 0.1589 |
| 2456714.2678 | 0.2087 | 0.0036 | 0.1526 | 2456714.2701 | -0.2279 | 0.0030 | 0.1422 | 2456714.2713 | -0.7007 | 0.0029 | 0.1365 |
| 2456714.2726 | 0.2765 | 0.0036 | 0.1304 | 2456714.2749 | -0.1580 | 0.0029 | 0.1200 | 2456714.2761 | -0.6292 | 0.0030 | 0.1144 |
| 2456714.2778 | 0.3628 | 0.0037 | 0.1065 | 2456714.2801 | -0.0669 | 0.0032 | 0.0961 | 2456714.2812 | -0.5372 | 0.0035 | 0.0909 |
| 2456714.2823 | 0.4644 | 0.0040 | 0.0860 | 2456714.2892 | 0.1783 | 0.0034 | 0.0544 | 2456714.2857 | -0.4198 | 0.0036 | 0.0702 |
| 2456714.2869 | 0.6029 | 0.0042 | 0.0649 | 2456714.2936 | 0.2955 | 0.0036 | 0.0340 | 2456714.2903 | -0.2937 | 0.0037 | 0.0493 |
| 2456714.2913 | 0.7346 | 0.0047 | 0.0445 | 2456714.2980 | 0.3145 | 0.0037 | 0.0137 | 2456714.2947 | -0.1962 | 0.0038 | 0.0289 |
| 2456714.2958 | 0.8024 | 0.0051 | 0.0241 | 2456714.3069 | 0.1700 | 0.0035 | 0.9727 | 2456714.2992 | -0.1905 | 0.0039 | 0.0085 |
| 2456714.3002 | 0.7976 | 0.0050 | 0.0036 | 2456714.3115 | 0.0311 | 0.0034 | 0.9516 | 2456714.3036 | -0.2324 | 0.0038 | 0.9881 |
| 2456714.3137 | 0.4296 | 0.0040 | 0.9417 | 2456714.3160 | -0.0778 | 0.0032 | 0.9313 | 2456714.3081 | -0.3523 | 0.0037 | 0.9675 |
| 2456714.3181 | 0.3413 | 0.0038 | 0.9213 | 2456714.3204 | -0.1580 | 0.0032 | 0.9109 | 2456714.3171 | -0.5745 | 0.0035 | 0.9262 |
| 2456714.3270 | 0.2096 | 0.0039 | 0.8806 | 2456714.3248 | -0.2068 | 0.0031 | 0.8904 | 2456714.3215 | -0.6470 | 0.0035 | 0.9058 |
| 2456714.3315 | 0.1786 | 0.0037 | 0.8600 | 2456714.3293 | -0.2559 | 0.0032 | 0.8701 | 2456714.3259 | -0.7080 | 0.0035 | 0.8854 |
| 2456714.3359 | 0.1462 | 0.0037 | 0.8394 | 2456714.3337 | -0.2919 | 0.0031 | 0.8495 | 2456714.3304 | -0.7461 | 0.0035 | 0.8649 |
| 2456714.3407 | 0.1144 | 0.0036 | 0.8174 | 2456714.3382 | -0.3176 | 0.0032 | 0.8289 | 2456714.3348 | -0.7778 | 0.0035 | 0.8444 |
| 2456714.3451 | 0.0963 | 0.0035 | 0.7971 | 2456714.3430 | -0.3434 | 0.0031 | 0.8070 | 2456714.3393 | -0.8059 | 0.0035 | 0.8238 |
| 2456714.3496 | 0.0916 | 0.0036 | 0.7766 | 2456714.3474 | -0.3529 | 0.0031 | 0.7866 | 2456714.3441 | -0.8210 | 0.0033 | 0.8019 |
| 2456714.3541 | 0.0990 | 0.0033 | 0.7561 | 2456714.3519 | -0.3574 | 0.0029 | 0.7661 | 2456714.3485 | -0.8372 | 0.0034 | 0.7814 |
| 2456714.3585 | 0.1043 | 0.0034 | 0.7355 | 2456714.3563 | -0.3559 | 0.0029 | 0.7457 | 2456714.3530 | -0.8349 | 0.0033 | 0.7609 |
| 2456714.3630 | 0.1206 | 0.0036 | 0.7150 | 2456714.3608 | -0.3333 | 0.0031 | 0.7250 | 2456714.3574 | -0.8259 | 0.0033 | 0.7405 |
| 2456714.3674 | 0.1520 | 0.0037 | 0.6946 | 2456714.3653 | -0.3127 | 0.0031 | 0.7045 | 2456714.3619 | -0.8141 | 0.0034 | 0.7199 |
| 2456714.3720 | 0.1801 | 0.0038 | 0.6738 | 2456714.3697 | -0.2797 | 0.0033 | 0.6840 | 2456714.3664 | -0.7859 | 0.0035 | 0.6995 |
| 2456714.3764 | 0.2160 | 0.0039 | 0.6534 | 2456714.3742 | -0.2606 | 0.0033 | 0.6634 | 2456714.3709 | -0.7655 | 0.0035 | 0.6787 |
| 2456714.3809 | 0.2712 | 0.0045 | 0.6327 | 2456714.3787 | -0.2042 | 0.0034 | 0.6428 | 2456714.3753 | -0.7243 | 0.0037 | 0.6583 |
| 2456714.3854 | 0.3371 | 0.0038 | 0.6122 | 2456714.3876 | -0.0765 | 0.0031 | 0.6018 | 2456714.3798 | -0.6876 | 0.0037 | 0.6376 |
| 2456714.3898 | 0.4549 | 0.0037 | 0.5918 | 2456714.3921 | 0.0313 | 0.0032 | 0.5811 | 2456714.3887 | -0.5462 | 0.0034 | 0.5966 |
| 2456714.3943 | 0.5748 | 0.0040 | 0.5712 | 2456714.3966 | 0.1647 | 0.0033 | 0.5607 | 2456714.3932 | -0.4354 | 0.0035 | 0.5760 |
| 2456714.3987 | 0.6996 | 0.0041 | 0.5508 | 2456714.4010 | 0.2744 | 0.0034 | 0.5404 | 2456714.3977 | -0.3090 | 0.0035 | 0.5557 |
| 2456714.4031 | 0.7477 | 0.0043 | 0.5304 | 2456714.4055 | 0.2913 | 0.0034 | 0.5198 | 2456714.4021 | -0.2307 | 0.0036 | 0.5353 |
| 2456714.4076 | 0.7493 | 0.0042 | 0.5098 | 2456714.4099 | 0.2871 | 0.0034 | 0.4993 | 2456714.4110 | -0.2289 | 0.0036 | 0.4943 |
| 2456714.4121 | 0.7171 | 0.0041 | 0.4894 | 2456714.4143 | 0.2247 | 0.0033 | 0.4789 | 2456714.4199 | -0.4413 | 0.0034 | 0.4532 |
| 2456714.4165 | 0.6098 | 0.0038 | 0.4690 | 2456714.4188 | 0.0902 | 0.0032 | 0.4584 | 2456714.4245 | -0.5544 | 0.0035 | 0.4323 |
| 2456714.4211 | 0.4765 | 0.0037 | 0.4479 | 2456714.4234 | -0.0264 | 0.0031 | 0.4375 | 2456714.4289 | -0.6406 | 0.0033 | 0.4119 |
| 2456714.4300 | 0.2782 | 0.0034 | 0.4069 | 2456714.4323 | -0.2055 | 0.0030 | 0.3964 | 2456714.4330 | -0.7036 | 0.0033 | 0.3913 |
| 2456714.4345 | 0.2246 | 0.0034 | 0.3864 | 2456714.4367 | -0.2302 | 0.0030 | 0.3760 | 2456714.4379 | -0.7415 | 0.0033 | 0.3708 |
| 2456714.4389 | 0.1863 | 0.0035 | 0.3660 | 2456714.4412 | -0.2700 | 0.0030 | 0.3555 | 2456714.4423 | -0.7672 | 0.0033 | 0.3504 |
| 2456714.4433 | 0.1571 | 0.0034 | 0.3456 | 2456714.4456 | -0.3039 | 0.0030 | 0.3350 | 2456714.4467 | -0.7943 | 0.0033 | 0.3300 |
| 2456714.4478 | 0.1324 | 0.0034 | 0.3251 | 2456714.4501 | -0.3348 | 0.0030 | 0.3146 | 2456714.4512 | -0.8234 | 0.0034 | 0.3095 |
| 2456714.4523 | 0.1083 | 0.0032 | 0.3046 | 2456714.4586 | -0.3543 | 0.0030 | 0.2755 | 2456714.4597 | -0.8391 | 0.0034 | 0.2705 |
| 2456714.4607 | 0.0863 | 0.0034 | 0.2656 | 2456714.4630 | -0.3704 | 0.0030 | 0.2551 | 2456714.4641 | -0.8260 | 0.0034 | 0.2500 |
| 2456714.4652 | 0.0988 | 0.0035 | 0.2451 | 2456714.4675 | -0.3523 | 0.0030 | 0.2347 | 2456714.4686 | -0.8280 | 0.0034 | 0.2296 |
| 2456714.4740 | 0.1302 | 0.0036 | 0.2045 | 2456714.4763 | -0.3150 | 0.0032 | 0.1941 | 2456714.4774 | -0.8009 | 0.0037 | 0.1889 |
| 2456714.4785 | 0.1567 | 0.0038 | 0.1840 | 2456714.4807 | -0.2904 | 0.0032 | 0.1737 | 2456714.4819 | -0.7681 | 0.0039 | 0.1685 |
| 2456714.4829 | 0.1880 | 0.0038 | 0.1636 | 2456714.4852 | -0.2578 | 0.0033 | 0.1533 | 2456714.4863 | -0.7304 | 0.0036 | 0.1482 |
| 2456714.4873 | 0.2388 | 0.0040 | 0.1433 | 2456714.4903 | -0.1900 | 0.0035 | 0.1299 | 2456714.4926 | -0.6337 | 0.0038 | 0.1190 |
| 2456714.4937 | 0.3336 | 0.0042 | 0.1142 | 2456714.4960 | -0.1019 | 0.0035 | 0.1037 | 2456714.4971 | -0.5608 | 0.0038 | 0.0987 |

Table 2. Light curve parameters for the eclipsing binary SWASP08.

| Filter | $D_{max}$ | $D_{min}$ | $A_p$ | $A_s$ |
|---|---|---|---|---|
| V | 0.0088±0.0004 | 0.0546±0.0022 | 0.7195±0.0294 | 0.6649±0.0271 |
| R | 0.0130±0.0005 | 0.0232±0.0010 | 0.6849±0.0280 | 0.6617±0.0270 |
| I | 0.0019±0.0001 | 0.0402±0.0016 | 0.6486±0.0265 | 0.6089±0.0249 |

Tables 3. New Times of minima of SWASP08.

| Min(I) | Error | Filter | Type | Ref. |
|---|---|---|---|---|
| 2456702.6603 | 0.0005 | V | I | Terrill and Gross (2014) |
| 2456702.7728 | 0.0004 | V | II | Terrill and Gross (2014) |
| 2456702.8819 | 0.0002 | V | I | Terrill and Gross (2014) |
| 2456703.6430 | 0.0001 | V | I | Terrill and Gross (2014) |
| 2456703.7517 | 0.0002 | V | II | Terrill and Gross (2014) |
| 2456703.8611 | 0.0011 | V | I | Terrill and Gross (2014) |
| 2456714.2973 | 0.0002 | V | I | This Paper |
| 2456714.4060 | 0.0002 | V | II | This Paper |
| 2456714.2978 | 0.0001 | R | I | This Paper |
| 2456714.4063 | 0.0003 | R | II | This Paper |
| 2456714.2979 | 0.0001 | I | I | This Paper |
| 2456714.4064 | 0.0005 | I | II | This Paper |

Table 4. Photometric solution for SWASP08 in VRI bands.

| Parameter | V filter | R filter | I Filter | VRI |
|---|---|---|---|---|
| A | 5500 | 7000 | 9000 | -- |
| $i(^0)$ | 86.89±0.94 | 86.79±0.83 | 86.16±0.95 | 82.70±0.27 |
| $g_1 = g_2$ | 0.32 | 0.32 | 0.32 | 0.32 |
| $A_1 = A_2$ | 0.5 | 0.5 | 0.5 | 0.5 |
| $q = (M_2/M_1)$ | 0.4800±0.0027 | 0.4815±0.0025 | 0.4738±0.0016 | 0.4732±0.0018 |
| $\Omega$ | 2.7516±0.0070 | 2.7668±0.0060 | 2.7497±0.0080 | 2.7641±0.0046 |
| $\Omega_{in}$ | 2.8488 | 2.8401 | 2.8251 | 2.8239 |
| $\Omega_{out}$ | 2.5578 | 2.5515 | 2.5407 | 2.5398 |
| $T_1^0 K$ | 4589 fixed | 4589 fixed | 4589 fixed | 4589 fixed |
| $T_2^0 K$ | 4561±4 | 4571±3 | 4569±5 | 4581±3 |
| $L_1/(L_1+L_2)$ | 0.6637 | 0.6609 | 0.6634 | --- |
| $L_2/(L_1+L_2)$ | 0.3344 | 0.3391 | 0.3366 | --- |
| $r_1$ pole | 0.4327±0.0031 | 0.4302±0.0029 | 0.4320±0.0032 | 0.4293±0.0027 |
| $r_1$ side | 0.4636±0.0043 | 0.4603±0.0039 | 0.4625±0.0044 | 0.4590±0.0036 |
| $r_1$ back | 0.4978±0.0061 | 0.4935±0.0055 | 0.4958±0.0062 | 0.4910±0.052 |
| $r_2$ pole | 0.3123±0.0041 | 0.3102±0.0038 | 0.3094±0.0043 | 0.3064±0.0039 |
| $r_2$ side | 0.3284±0.0052 | 0.3258±0.0047 | 0.3251±0.0053 | 0.3213±0.0048 |
| $r_2$ back | 0.3728±0.0097 | 0.3681±0.0086 | 0.3679±0.0098 | 0.3617±0.0088 |
| $\Sigma$ (O-C) | 0.0018 | 0.0016 | 0.0019 | 0.0073 |

Table 5. Absolute physical parameters for the system for the system SWASP08.

| Element | M ($M_\odot$) | R ($R_\odot$) | T ($T_\odot$) | $L_1$ ($L_\odot$) | $M_{bol}$ | Sp. Type. | Dist (pc) |
|---|---|---|---|---|---|---|---|
| Primary | 0.70±0.29 | 0.80±0.03 | 0.80±0.03 | 0.25±0.01 | 6.27±0.26 | K4 | 195 |
| Secondary | 0.34±0.01 | 0.78±0.03 | 0.80±0.03 | 0.23±0.01 | 6.32±0.26 | | |